\documentclass[twocolumn]{aastex63}

\usepackage{textcomp}
\usepackage{bm,color}
\usepackage{verbatim}
\usepackage{amsmath}
\usepackage{ulem}
\usepackage{hyperref}  
\usepackage{booktabs}
\usepackage{bigstrut}
\usepackage{xspace}
\usepackage{multirow}
\usepackage{amsmath}
\usepackage{graphicx}
\usepackage{threeparttable}

\def \BE{\begin{equation}}
\def \EE{\end{equation}}	
\def \BC{\begin{center}}
\def \EC{\end{center}}
\def \BEA{\begin{eqnarray}}
\def \EEA{\end{eqnarray}}

\def \SIGMA8{\sigma_{8}}

\def\lcdm{\ensuremath{\Lambda {\rm CDM}}}

\newcommand{\skipt}[1]{}

\defcitealias{gupta20a}{G20}

\begin{document}\sloppy\sloppypar\raggedbottom\frenchspacing

\title[Deep learning: Galaxy cluster mass estimates]{Mass Estimation of Galaxy Clusters with Deep Learning II: CMB Cluster Lensing }

\shortauthors{N.~Gupta, et al.}
\author[0000-0001-7652-9451]{N.~Gupta} \thanks{Nikhel.Gupta@csiro.au} \affiliation{School of Physics, University of Melbourne, Parkville, VIC 3010, Australia}
\affiliation{CSIRO Space \& Astronomy, PO Box 1130, Bentley WA 6102, Australia} 
\author[0000-0003-2226-9169]{C.~L.~Reichardt} \affiliation{School of Physics, University of Melbourne, Parkville, VIC 3010, Australia}

\begin{abstract}
We present a new application of deep learning to reconstruct the cosmic microwave background (CMB) temperature maps from the images of microwave sky, and to use these reconstructed maps to estimate the masses of galaxy clusters.
We use a feed-forward deep learning network, mResUNet, for both steps of the analysis. 
The first deep learning model, mResUNet-I, is trained to reconstruct foreground and noise suppressed CMB maps from a set of simulated images of the microwave sky that include signals from the cosmic microwave background, astrophysical foregrounds like dusty and radio galaxies, instrumental noise as well as the cluster's own thermal Sunyaev Zel'dovich  signal. 
The second deep learning model, mResUNet-II, is trained to estimate cluster masses from the gravitational lensing signature in the reconstructed foreground and noise suppressed CMB maps. 
For SPTpol-like noise levels, the trained mResUNet-II model recovers the mass for $10^4$ galaxy cluster samples with a 1-$\sigma$ uncertainty $\Delta M_{\rm 200c}^{\rm est}/M_{\rm 200c}^{\rm est} =$ 0.108 and 0.016 for input cluster mass $M_{\rm 200c}^{\rm true}=10^{14}~\rm M_{\odot}$ and $8\times 10^{14}~\rm M_{\odot}$, respectively. 
We also test for potential bias on recovered masses, finding that for a set of $10^5$ clusters the estimator recovers $M_{\rm 200c}^{\rm est} = 2.02 \times 10^{14}~\rm M_{\odot}$, consistent with the input at 1\% level. The 2 $\sigma$ upper limit on potential bias is at 3.5\% level.
\end{abstract}


\section{Introduction} 
\label{sec:intro}
The number density of galaxy clusters is a promising approach to constrain cosmological models, especially those affecting late-time structure growth \citep[e.g.][]{mantz08, vikhlinin09, hasselfield13, planck16-24, dehaan16, bocquet19, constanzi19}.
Current and upcoming experiments are expected to collectively detect more than $10^5$ galaxy clusters in the next few years, two orders of magnitude larger than current cluster catalogs. 
Data is already being collected by some of these experiments \citep[e.g. {\it eROSITA}, SPT-3G, AdvancedACT:][]{ predehl10, benson14, henderson16} and others plan to start operations in a near future \citep[e.g. LSST, Euclid, Simons Observatory, CMB-S4:][]{lsst09, laurejis11, simons19, cmbs4-19}.
The dramatically larger cluster catalogs have the potential to have a huge impact on our understanding of structure growth and the expansion history of universe.

One major hurdle for extracting full cosmological information from these galaxy cluster catalogs is the calibration between the cluster mass and observables \citep[see, e.g.][]{bocquet15, Planck15, constanzi20}. 
Several techniques have been used to estimate the mass of galaxy clusters. 
Among these, the weak gravitational lensing of background galaxies \citep[e.g.][]{johnston07, gruen14, hoekstra15, stern18, mclintock19} and the lensing of cosmic microwave background \citep[CMB, e.g.][]{seljak00b,dodelson04,holder04,maturi05,lewis06,hu07,yoo08,baxter15,melin15, madhavacheril15, planck16-24,geach17b,baxter18, madhavacheril18, raghunathan19} by galaxy clusters, have demonstrated the potential for unbiased mass measurements.

Optical weak lensing becomes difficult for high-redshift galaxy clusters due to the decreasing density of background galaxies. In contrast, CMB lensing works well for at high redshift as CMB originates at $z\sim 1100$.
However, CMB lensing has its own limitations related to the raw signal-to-noise and the presence of foregrounds in CMB  maps.
A significant foreground for CMB lensing is the thermal Sunyaev-Zel'dovich (tSZ) effect \citep[][]{sunyaev70, sunyaev72} signal of cluster itself that arises due to the inverse-Compton scattering of the CMB photons by energetic electrons in intra cluster medium (ICM). 
If not handled, the tSZ signal (and any other millimeter-wave signals sourced by the galaxy cluster) will bias the CMB lensing mass measurement of a galaxy cluster. 

In the standard quadratic estimator (QE), the large scale CMB gradient and the small scale CMB anisotropy maps are used, and the lensing signal is extracted using the correlation between the different angular scales that are uncorrelated in the primordial CMB anisotropy map \citep{hu07}.
Leveraging on the frequency dependence of the tSZ signal, the gradient CMB map can be freed from the tSZ that eliminates the induced correlation with the CMB anisotropy map  \citep{madhavacheril18, raghunathan19}. 
While this approach eliminates the bias in gradient maps, the tSZ power in small scale CMB anisotropy map adds extra variance that can become significant for high mass clusters and low noise surveys.
Few other methods are devised to reduce the lensing bias and variance due to the tSZ signal from galaxy clusters.
These methods include the inpainting of gradient map based on the information from surrounding pixels \citep{raghunathan19b} and tSZ template fitting \citep{patil20}. 

The maximum likelihood estimator (MLE) on the other hand extracts optimal lensing signal from CMB temperature maps \citep[e.g.][]{dodelson04,baxter15,raghunathan17}.
The approach is based on fitting the lensed CMB templates to observed CMB maps and performs better than the standard QE by a factor of two at very low noise levels in the absence of tSZ signal and astrophysical foregrounds. 
However, current MLEs depend on the map being clean of foregrounds and are biased when applied to maps with residual foregrounds \citep{raghunathan17}. 

In this work, we demonstrate the first use of a deep learning network to estimate the mass of galaxy clusters from the CMB lensing signal.
We employ a modified version of a feed-forward deep learning algorithm, mResUNet \citep[][hereafter G20]{gupta20a} that combines residual learning \citep{he15d} and U-Net framework \citep{ronneberger15d}.
Two separate mResUNet models are trained independently. 
In the first step, we reconstruct the CMB temperature maps  from the simulated images of microwave sky maps.
This is done by training the mResUNet-I network to learn CMB features and mitigate the foreground (tSZ and astrophysical) signals as well as instrumental noise.
In the second step, we use the reconstructed CMB temperature maps and mResUNet-II network to estimate the underlying mass for individual galaxy clusters. 
The mResUNet-II network is trained to extract lensing features from CMB temperature maps.
After training these models for CMB reconstruction and mass estimation, we test the robustness of the process by using external hydrodynamical simulations of galaxy clusters.

The paper is structured as follows. In Section~\ref{sec:methods}, we describe the simulations of microwave sky maps, the deep learning model and the parameters for its optimization.
In Section~\ref{sec:results}, we present mass predictions using the images from test data sets as well as the images from the external hydrodynamical simulations of SZ clusters.
In Section~\ref{sec:conclusions}, we summarize our findings and discuss future prospects.

Throughout this paper, $M_{\rm 200c}$ is defined as the mass of the cluster within the region where the average mass density is 200 times the critical density of universe. 
Unless specified, the term `foregrounds' refers to both the tSZ signal and astrophysical emission. 
The central mass and the 1-$\sigma$ uncertainty is calculated as the median and the standard deviation divided by $\sqrt{N}$ (standard error), where $N$ is the number of clusters, respectively.

\section{Methods}
\label{sec:methods}
In this section, we describe the microwave sky simulations of the lensed CMB temperature anisotropy; tSZ effect; radio and dusty galaxies; and instrumental noise. 
We then discuss the deep learning models used to extract foreground-cleaned CMB temperature maps and to estimate the masses of galaxy clusters.

\subsection{Simulations of Lensed CMB Temperature Maps}
\label{sec:SZE}

We create sets of simulations at each of fifteen cluster masses, ranging from 1 - $8\times 10^{14}~\rm M_{\odot}$ in steps of $0.5\times 10^{14}~\rm M_{\odot}$. 
All clusters are assumed to lie at $z=0.7$. 
These simulations include the CMB (lensed by the cluster), astrophysical foregrounds (not lensed), the effects of the instrumental beam, and instrumental noise.
The simulated maps are trimmed to a box size of $10^{\prime} \times 10^{\prime}$ with a  pixel resolution of $0.25^{\prime}$. 
A Hanning window is applied to the outer $2^{\prime}$ of the box to prepare for the convolutions in the neural network. 

The first ingredient in these simulated maps is the CMB. 
We generate Gaussian realizations of the best-fit lensed \lcdm{} cosmology for the {\it Planck} 2016 cosmology results \citet{planck16-CMB}; the power spectrum is calculated using \texttt{CAMB}\footnote{\url{https://camb.info/}} \citep{lewis00}. 
As we do for all the fields, we generate these CMB realizations on a larger box ($60^{\prime} \times 60^{\prime}$) to avoid edge effects from the Fourier transforms and to be sure we capture the large-scale gradients across the final box. 
The CMB maps are then lensed by the galaxy cluster at their center. 
We use interpolation to handle the sub-pixel-scale deflection angles. 
The cluster lensing convergence profile, $\kappa(M,z)$, is  based on the projected NFW profile given by \citep{bartelmann96}. 

To these cluster-lensed CMB maps, we add both astrophysical foregrounds unassociated with the cluster and the cluster's own SZ signal. 
For the first, we create Gaussian realizations of the best-fit baseline foreground model in \citet{george15} at 150\,GHz. 
This model includes power from other SZ halos, the kinematic SZ effect, radio galaxies, and the cosmic infrared background (CIB). 
For the second, we follow \citetalias{gupta20a} and add a tSZ template at the cluster position based on the GNFW profile. 
We scale the amplitude and size of this profile following the power law relations between the tSZ signal, mass and redshift given in \citet{arnaud10}. 
We also add a 20\% log-normal scatter in the peak amplitude of the tSZ profile. 
Neither of the foregrounds nor cluster's own SZ signal are lensed by the cluster. 

We then convolve both the cluster-lensed CMB and foreground maps by a Gaussian instrumental beam with FWHM $=1^{\prime}$. 
We store three combinations of these maps for each realization. 
The first  combination map is representative of what a telescope would see; it has the sum of the CMB, foregrounds, and a realization of $5~\rm \mu K$-arcmin white noise.
We find that the network is not biased by modest changes in the map noise. 
Specifically, we find no bias in the recovered mass when providing maps that have  30\% higher white noise to a network trained with baseline noise. The test was done for a sample of 10,000 clusters with $M_{\rm 200c} = 4\times 10^{14}~\rm M_{\odot}$, with a 1 $\sigma$ uncertainty on the recovered mass of 3.8\%.
Note that these simulations neglect correlated instrumental noise, as well as non-Gaussianity in the astrophysical foregrounds. 
Future work should assess the impact of these on the deep learning estimator.
The second combination map is used as a `truth' set in training the first network; it has only the CMB.
The third combination map is the input for training the second network; this map has the CMB and a realization of white noise.
We randomly draw the white noise level for this third map from a uniform distribution between 0 to $5~\rm \mu K$-arcmin.
Varying noise levels is designed to diversify the training set to better handle the non-Gaussian statistics in the output map from stage I (see Fig.~\ref{FIG:T_residual}). 
We do not use the maps from stage I for training the second stage in order to ensure that the network is not trained to recover masses using information besides lensing, for example from some residual SZ signal. 
The $\kappa$ maps are used as the `truth' sets in training the second network. 
All maps are then trimmed to $10^{\prime} \times 10^{\prime}$ and apodized as described above. 
We have tested training the networks after trimming the maps to larger ($20^{\prime} \times 20^{\prime}$) sizes at a single cluster mass; we found a small improvement for $20^{\prime} \times 20^{\prime}$ maps. 
The uncertainty decreased by a factor of 0.93 to 0.98 for $20^{\prime} \times 20^{\prime}$ maps, but consumed more computer resources.
All results shown in this work use $10^{\prime} \times 10^{\prime}$ maps.

\begin{figure*}
\centering
\includegraphics[trim={2cm 0.5cm 2cm 3cm},clip,width=18cm, scale=0.5]{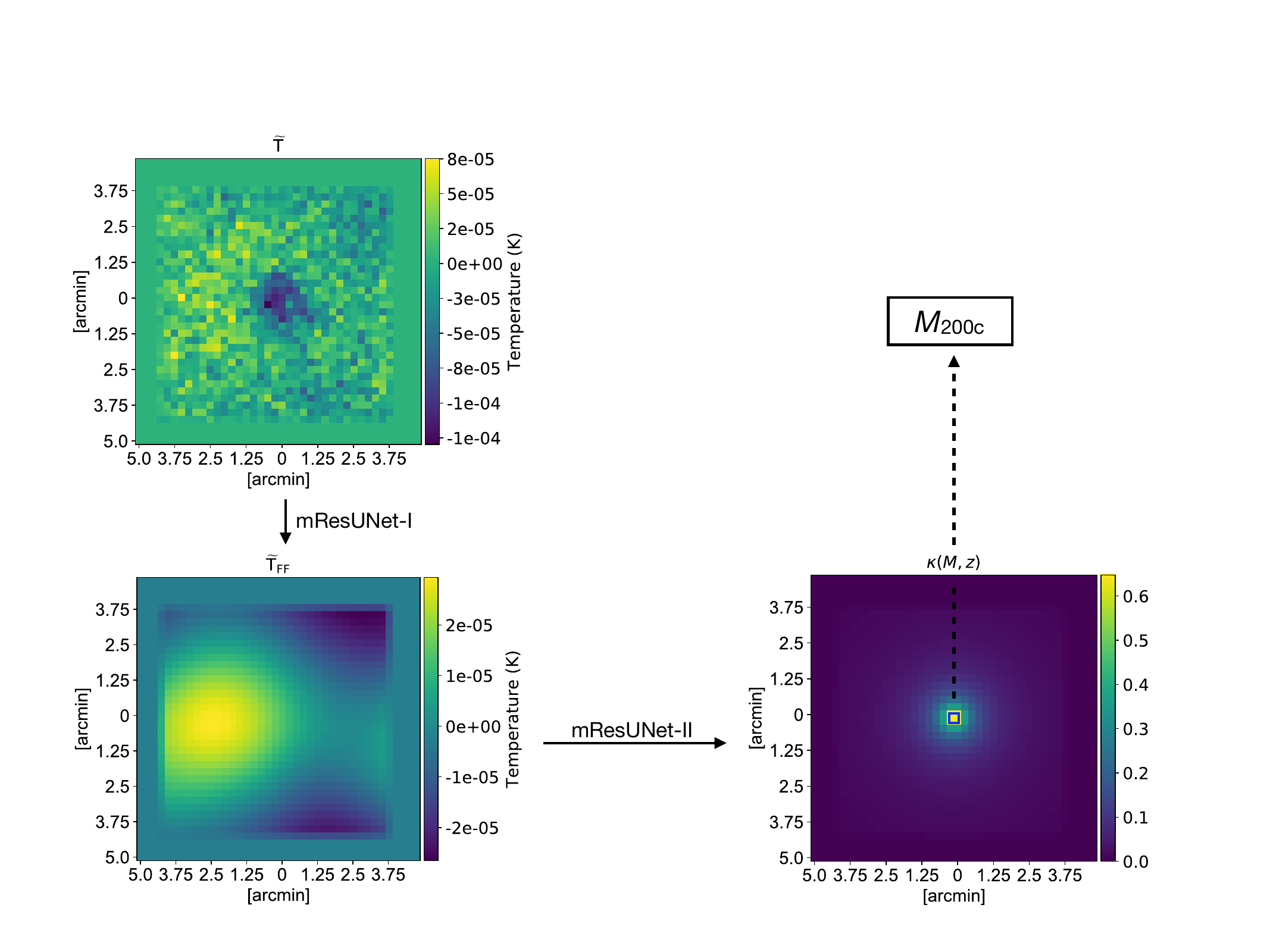}
\caption{The work flow in the analysis: Top left panel shows an example of microwave sky CMB map ($\rm \widetilde{T}$, where tilde represents lensing) for a cluster with $M_{\rm 200c} = 4\times 10^{14}~\rm M_{\odot}$ at $z=0.7$. 
This map includes cluster-lensed CMB, astrophysical foregrounds, cluster tSZ signal and an instrumental white noise of $5~\rm \mu K$-arcmin.
The map is convolved with $1^{\prime}$ telescope beam and apodization is applied.
Several such maps for different cluster masses are used for training of the neural network (mResUNet-I) to recover foreground and noise suppressed maps ($\rm \widetilde{T}_{\rm FF}$) in the first stage of the analysis.
In the second stage of the analysis, the cluster-lensed CMB maps with varying white noise levels are  used to train an another network (mResUNet-II) to extract the lensing $\kappa(M,z)$ maps. The trained network is then applied to the foreground-cleaned CMB maps that are output by the first network. 
As described in Section~\ref{sec:SZE}, the mass of cluster is then estimated from the central pixel of the extracted $\kappa(M,z)$ map.}
\label{FIG:simulation}
\end{figure*}

We end up with four maps per realization: the sky map with the CMB, foregrounds and $5~\rm \mu K$-arcmin noise; the CMB-only map; the CMB map with a noise level uniformly drawn from the range 0 to $5~\rm \mu K$-arcmin; and the lensing convergence $\kappa$ map.
Since we are interested in the cluster mass only, we replace the central pixel of  each $\kappa(M,z)$ map, such that, the product of the central pixel value and the mean mass of the training sample equals $M_{\rm 200c}$ of the cluster.
The first network is trained on the sky maps to reconstruct the CMB-only maps. 
The second network is trained on the CMB maps with noise between 0 to $5~\rm \mu K$-arcmin level to reproduce the $\kappa$ maps and recover the cluster mass.
Figure~\ref{FIG:simulation} shows the examples of the sky map ($\rm \widetilde{T}$), reconstructed CMB-only map ($\rm \widetilde{T}_{\rm FF}$) and lensing convergence $\kappa(M,z)$ map for a cluster with $M_{200\rm c} =4\times 10^{14}~\rm M_{\odot}$.

\subsection{The mResUNet Model}
\label{sec:deep_learning}
We employ mResUNet \citep{caldeira19d}, a feed-forward Convolutional Neural Network (CNN), in the two legs of the analysis.
The modified ResUNet, or mResUNet, algorithm was adapted by \citet{caldeira19d} to do image-to-image regression, i.e. get an output image that is a continuous function of the input image, from the original ResUNet feed-forward deep learning algorithm \citep{kayalibay17, zhang18d}. 
The mResUNet, algorithm is well suited to astrophysical problems, such as the current use case of estimating the lensed CMB signal from an image of the sky. 

\citetalias{gupta20a} tuned the mResUnet network by introducing dilation rates to extract small and large scale features in a CMB map, which we also use in this work.
As shown in Figure~1 of \citetalias{gupta20a}, the framework is based on the encoder-decoder paradigm. 
This consists of a contracting path (encoder) to capture features, a symmetric expanding path (decoder) that enables precise localization.
Each path has several convolution blocks and each of these blocks have four sub-stages. 
Every sub-stage has a convolution layer, an activation function and a batch normalization layer.
The aim of the convolution layer is to learn features of an input map using filters that are applied to a receptive field of neighbouring pixels. 
Each filter is typically a $k \times k$ array with $k=1,3,5, ...$, and the size of the filter ($k \times k$) is denoted as the kernel size.
For each of the four sub-stages, we apply dilations to the convolution layers with a rate of 1, 2, 3 and 4. 
A dilation rate of N stretches the receptive field by $k + (k-1)(N-1)$, thus doubling the dilation rate from 1 to 2 increases the receptive field from $3\times3$ to $5 \times 5$ for $k$=3.
These dilated convolutions systematically aggregate multi-scale contextual information without losing resolution \citep{yu15d}.
The total receptive field increases for each pixel of the input image as we stack several convolution layers in the network.
An activation function is applied after each convolution layer in order to detect non-linear features, leading to a highly non-linear reconstruction of input image \citep[see][for a recent review]{nwankpa18d}. 
The batch normalization layer is helpful in improving the speed, stability and performance of the network.
The input to each convolution block is always added to its output using residual connections.
To avoid overfitting, we add dropout layers to the decoding phase of the network.
The weights of the network are optimized using gradient descent \citep[e.g.][]{ruder16d} that involves back-propagation from the final output, back to each layer in reverse order to update the weights.

Figure~\ref{FIG:simulation} shows the overview of our analysis.
In the first stage of analysis, we train the mResUNet-I network to reconstruct the CMB-only maps from the input sky maps that have CMB, noise and foregrounds. 
In the second stage, we train the mResUNet-II network to estimate the lensing convergence map, and specifically the mass of galaxy clusters, from the recovered CMB-only maps.
\begin{figure*}
\centering
\includegraphics[width=18cm, scale=0.5]{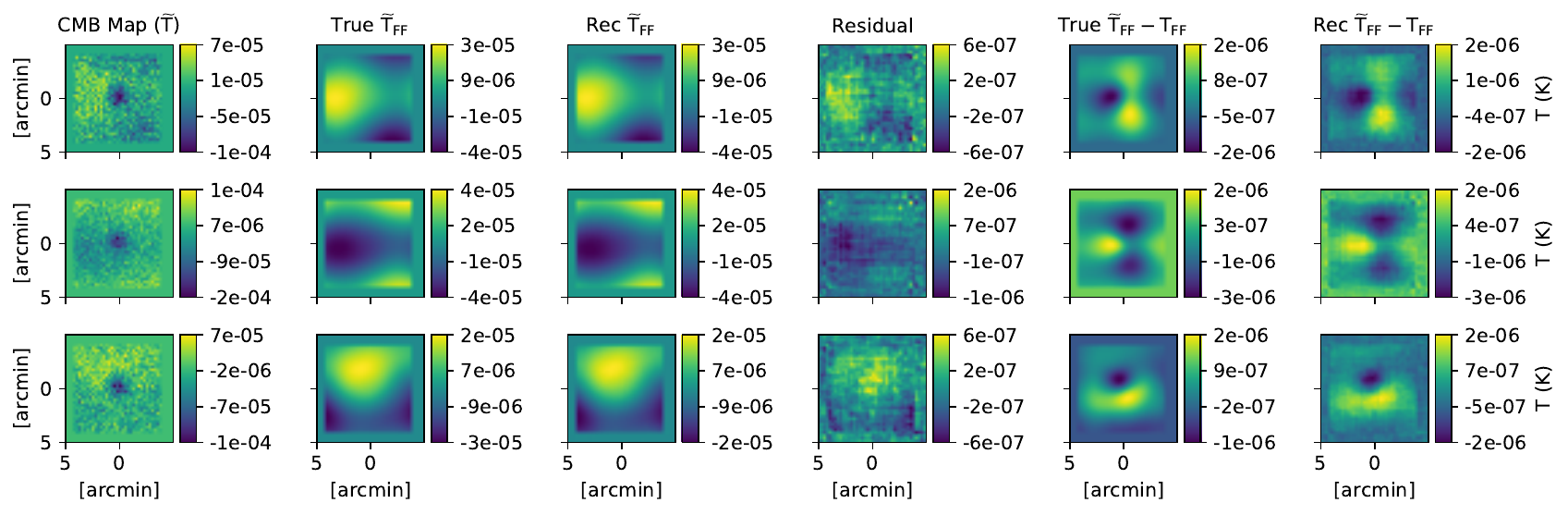}
\caption{\label{FIG:SZEprofile}
This figure illustrates the performance of the method for a random set of three galaxy clusters, with $M_{200\rm c} = 4 \times 10^{14}~\rm M_{\odot}$. 
Each row focuses on single galaxy cluster, with the columns showing maps from different stages in the analysis. 
The first column shows the microwave sky ($\rm \widetilde{T}$), including the cluster-lensed CMB, foregrounds and instrumental noise. 
The second and third columns show the true and recovered CMB-only maps ($\rm \widetilde{T}_{\rm FF}$). 
The difference between these two is shown in the fourth column, labelled `Residual'. 
The last two columns demonstrate the recovery of the desired  CMB-cluster lensing signal, by showing first the true lensed - unlensed CMB map in column 5, and second the recovered lensed CMB map - true unlensed CMB map in column 6.  
The trained mResUNet-I model successfully reconstructs the CMB map at high accuracy including the critical CMB-cluster lensing signal.
}
\end{figure*}
\subsection{Training and Optimisation}
\label{sec:optimisation}
The mResUNet-I and the mResUNet-II models take images as input and output same sized images after passing through several convolutional blocks.
The mResUNet-I network is trained to reconstruct $\rm \widetilde{T}_{\rm FF}$ maps and the mResUNet-II network is trained to extract the $\kappa(M,z)$ maps.
The central pixels of the extracted $\kappa(M,z)$ maps are used to estimate the mass of clusters (see Section~\ref{sec:SZE}).
Note that a different image-to-float regression method can be used in the second step instead of the image-to-image regression model mResUNet-II.
For simplicity, we chose the same network structure in both steps.
A future study should compare different deep learning networks to identify the most suitable model.

For training these networks, we normalize all input maps, so that the minimum and the maximum pixel value is between -1 and 1, respectively.
This is done by dividing the image pixels by a constant factor across all cluster masses. 
The data for both networks are divided into three parts: training, validation and test sets.
The training and the validation datasets are used in the model learning process.
In both stages of the analysis, the training data has 400 maps and corresponding truth sets for each cluster.
For first model, we take cluster simulations with $M_{200\rm c} =$ (1, 2, 3, 4, 5, 6, 7, 8)$\times 10^{14}~\rm M_{\odot}$ for training.
In second stage, the model is trained with $M_{200\rm c} =~\pm$ (1, 2, 3, 4, 5, 6, 7, 8, 9, 10, 20...90, 100, 200...500)$\times 10^{14}~\rm M_{\odot}$. 
We find that $M_{200\rm c} = \pm ~5 \times 10^{16}~\rm M_{\odot}$ is the optimized mass range for training the second model and for a narrower range, model underestimates the mass uncertainties for clusters in the mass range of first model.

The validation set has same properties as the training set and is used to validate the model after each epoch, where one epoch is complete when entire training data are passed through the neural network once. 
We use 200 maps for each cluster mass and corresponding truth sets as our validation data. 
The test datasets are never used in the training phase and are kept separately to analyse the trained model.
We keep $10^4$ sky maps (for each cluster mass) for testing in the first stage of analysis and use recovered CMB-only maps from these test data to estimate the mass of the individual clusters in the second stage of analysis. 

We use CMB maps with random realizations of white noise drawn from a uniform distribution between 0 to $5~\rm \mu K$-arcmin level to train (400 maps) and validate (200 maps) the network in the second stage of analysis.
In addition to the mass set used in training, we test our mResUNet-I and mResUNet-II models for cluster masses that were not the part of training and validation process, that is clusters with $M_{200\rm c} =$ (1.5, 2.5, 3.5, 4.5, 5.5, 6.5, 7.5)$\times 10^{14}~\rm M_{\odot}$.

The maps in the training set are passed through the neural networks with a batch size of 4 (16) for the first (second) model and a training loss is computed as mean-squared-error (MSE) between the predicted and the true labels after each batch.
Batch after batch, the weights of the network are updated using the gradient descent and the back-propagation.
In this work, we use Adam optimizer \citep[an algorithm for first-order gradient-based optimization, see][]{kingma14d} with an initial learning rate of 0.001. 
After each epoch, the validation loss (or validation MSE) is calculated and we change the learning rate by implementing callbacks during the training, such that the learning rate is reduced to half if the validation loss does not improve for five consecutive epochs.
In addition, to avoid over-fitting, we set a dropout rate of 0.3 in the encoding phase of the mResUNet-I network which is reduced to 0.2 for the mResUNet-II network.
We consider the network to be trained and stop the training process, if the validation loss does not improve for fifteen epochs.

As described in Section~\ref{sec:deep_learning}, each sub-stage of a convolution block in both models, has a convolution layer, an activation layer and a batch normalization layer.
We set the kernel-size of each convolution layer to $3\times 3$ and change the stride length from 1 to 2, whenever the filter size is doubled.
All activation layers in these networks have a Scale Exponential Linear Unit \citep[SELU][]{klambauer17d} activation functions which induce self-normalizing properties, such that, activations close to zero mean and unit variance converge towards zero mean and unit variance, when propagated through many network layers, even under the presence of noise and perturbations.
Only for the final layer in both models, a linear (or identity) activation function is used to get same sized output images as inputs.
Each network has approximately 16 million parameters and is separately trained on a single GPU using Keras with a TensorFlow backend.
For first and second network, training took roughly 100 and 240 seconds per epoch, respectively for a total of ten hours.
The trained networks take a total of about thirty minutes to run over all 150,000 realizations in the test set including the time to load the network and simulations from disk.
\begin{figure}
\centering
\includegraphics[width=8.4cm, scale=0.5]{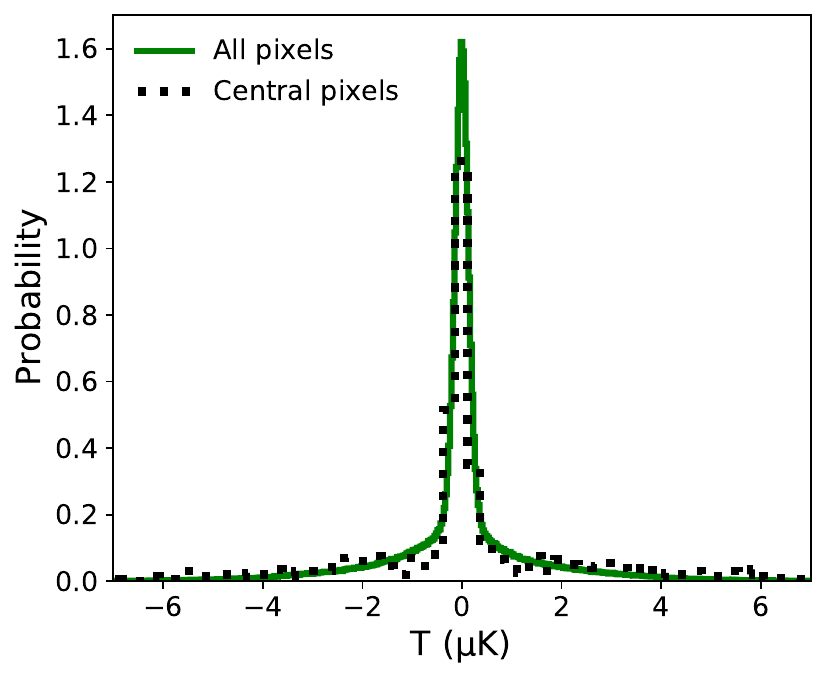}
\caption{The distribution of the residual temperature that is the difference between the true and the recovered CMB-only maps ($ \rm \widetilde{T}_{\rm FF}$) as shown in the fourth column of Figure~\ref{FIG:SZEprofile}. The green solid and the black dotted contours show the distribution of all image pixels and central image pixels for 800 realizations of a cluster with $M_{200\rm c} = 4 \times 10^{14}~\rm M_{\odot}$, respectively. 
The distribution is clearly non-Gaussian, resembling  a Lorentz distribution with $\gamma=0.5$. 
This indicates that the noise in the recovered $ \rm \widetilde{T}_{\rm FF}$ maps is non-Gaussian.}
\label{FIG:T_residual}
\end{figure}
\section{Results}
\label{sec:results}
In this section, we look at the performance of our trained neural network models.
We show the reconstructed CMB-only maps with foreground and noise suppressed ($\rm \widetilde{T}_{\rm FF}$) using the mResUNet-I network. 
We demonstrate that the mass of galaxy clusters can be estimated directly from the reconstructed $\rm \widetilde{T}_{\rm FF}$ maps using the mResUNet-II network. 
We find per cluster $\Delta M_{\rm 200c}^{\rm est}/M_{\rm 200c}^{\rm est} =$ 10.8 and 1.6 for input cluster mass $M_{\rm 200c}^{\rm true}=10^{14}~\rm M_{\odot}$ and $8\times 10^{14}~\rm M_{\odot}$, respectively.
We test the performance of our trained deep learning models by applying them to the sky maps with more realistic tSZ signal from the external hydrodynamical simulations.
Finally, we compare the $S/N$ of the mass estimations from the current analysis to those estimated with the MLE. 

\subsection{Reconstruction of $\rm \widetilde{T}_{\rm FF}$ Maps with mResUNet-I}
\label{sec:predictions1}
We use the test dataset of $10^4$ sky maps for each of the clusters to recover the CMB-only maps using the trained mResUNet-I model.
These test maps are not used for training and validation purposes and are distinct due to the Gaussian random realizations of the CMB and foregrounds as well as the 20\% log-normal scatter in the estimation of the tSZ signal (see Section~\ref{sec:SZE}).
The first column in Figure~\ref{FIG:SZEprofile} shows the examples of the input sky maps for three random realizations of a cluster with $M_{200\rm c} = 4\times 10^{14}~\rm M_{\odot}$.
The second and the third columns show the true and the reconstructed CMB-only maps, respectively.
The fourth column shows residual signals, that is the difference between the true and the recovered CMB-only maps.
The fifth column shows the difference between the true lensed and the unlensed CMB-only maps, and the last column shows the same for the recovered CMB-only maps.
This demonstrates that the trained mResUNet-I model successfully reconstructs the CMB map at high accuracy including the critical CMB-cluster lensing signal.

Figure~\ref{FIG:T_residual} shows the distributions of the residual temperature calculated as the difference between the true and the recovered CMB-only maps. 
The green solid and black dotted contours show the distribution of all image pixels and central pixels, respectively, for 800 realizations of the cluster with $M_{200\rm c} = 4\times 10^{14}~\rm M_{\odot}$. 
This indicates that the noise in the recovered CMB-only maps is non-Gaussian and resembles a Lorentz distribution with $\gamma=0.5$.

\begin{figure}
\centering
\includegraphics[width=8.4cm, scale=0.5]{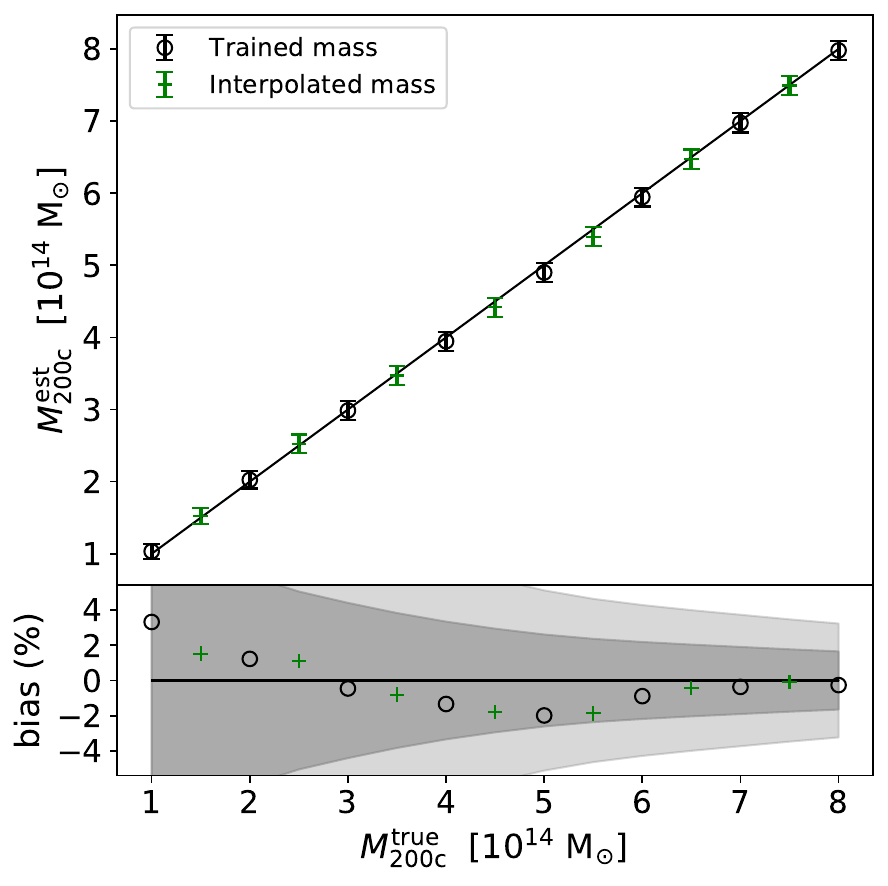}
\caption{The mass estimates from the trained mResUNet-II model. 
The top panel shows the estimated and the true mass of clusters using a test data set of $10^4$ CMB-only maps (with foreground and noise suppressed) per cluster mass that are recovered from the microwave sky maps in the first stage of analysis. 
The uncertainties are for $10^4$ clusters of a given mass.
The black data points show results for cluster masses equal to one of the training sets.
The green data points (interpolation) describe results for clusters with masses between the trained masses.
The bottom panel shows the percentage bias between the median estimated mass and true mass for a set of $10^4$ clusters.
The dark and light shaded regions denote the  1 and 2 $\sigma$ standard error.
This shows that the mass estimations from the deep learning method are unbiased.
For instance, the bias for cluster $M_{200 \rm c}=4 \times 10^{14}~\rm M_{\odot}$ is $< 1.5$\% with a 1 $\sigma$ limit at $3.5\%$ level.
}
\label{FIG:mass}
\end{figure}
\subsection{Mass Estimation with mResUNet-II}
\label{sec:predictions2}
We estimate the mass of galaxy clusters with the trained mResUNet-II network and using the $10^4$ reconstructed CMB-only maps from the first stage of analysis.
As the second network is trained on the cluster-lensed CMB maps with random noise realizations drawn from the uniform distribution between 0 to $5~\rm \mu K$-arcmin level, it extracts the $\kappa$ maps using only the cluster lensing features in the reconstructed CMB-only maps.
The mass of galaxy clusters is then estimated from the $10^4$ extracted $\kappa$ maps.
As described in Section~\ref{sec:SZE}, this is done by multiplying the central pixel of the extracted $\kappa$ maps by the mean mass of the training sample.
The top panel in Figure~\ref{FIG:mass} shows the estimated mass of single clusters as a function of the true mass.
The black data points show estimations for cluster masses that are trained in the network and the green data points describe the estimations for the interpolated mass set.
All estimations are in the range of trained mass sample ($1\times 10^{14}~\rm M_{\odot} < M_{200\rm c} < 8\times 10^{14}~\rm M_{\odot}$).
This shows that our trained mResUNet-II model can estimate the mass of  with good precision.
For instance, we find $M_{200 \rm c}^{\rm est}=3.94\pm0.13 \times 10^{14}~\rm M_{\odot}$ for $10^4$ clusters with $M_{200 \rm c}^{\rm true}=4 \times 10^{14}~\rm M_{\odot}$.
While we have not looked at extrapolated masses in this work, we expect the network to show the same behaviour as in \citetalias{gupta20a}. As shown in Figure 4 of that work, the network returned the highest (lowest) trained mass for clusters that were above (below) that mass. The training sample should encompass the entire cluster mass range of interest.

We estimate the bias for each mass as
\BE
\label{EQ:bias}
{\rm bias} =\frac{ M_{200 \rm c}^{\rm est}}{ M_{200 \rm c}^{\rm true}} -1,
\EE
and find that the bias in the median mass estimation from this deep learning method is $\leq3\%$ and within the 1 $\sigma$ level of standard error for all masses (see bottom panel of Fig.~\ref{FIG:mass}).
For instance, the bias for cluster mass $4 \times 10^{14}~\rm M_{\odot}$ is $< 1.5\%$ with a 1 $\sigma$  limit for $10^4$ clusters at $3.5\%$ level.
This shows no evidence of bias from the estimator.
While reassuring, we also confirm the non-detection for a lower mass cluster with $M_{200 \rm c}^{\rm true}= 2 \times 10^{14}~\rm M_{\odot}$ using $10^5$ realizations, finding $1\%$ bias with a 2 $\sigma$ upper limit at 3.5\% level. 
This is much smaller than the 6.3\% bias estimated by \cite{raghunathan17} (see Table 1 and section 4.6.2 in their paper) using MLE with 1\% residual tSZ signal in maps, indicating that the first step our deep learning method possibly leaves $< 1\%$ residual tSZ signal in the reconstructed maps.

\begin{figure}
\centering
\includegraphics[width=8.2cm, scale=0.5]{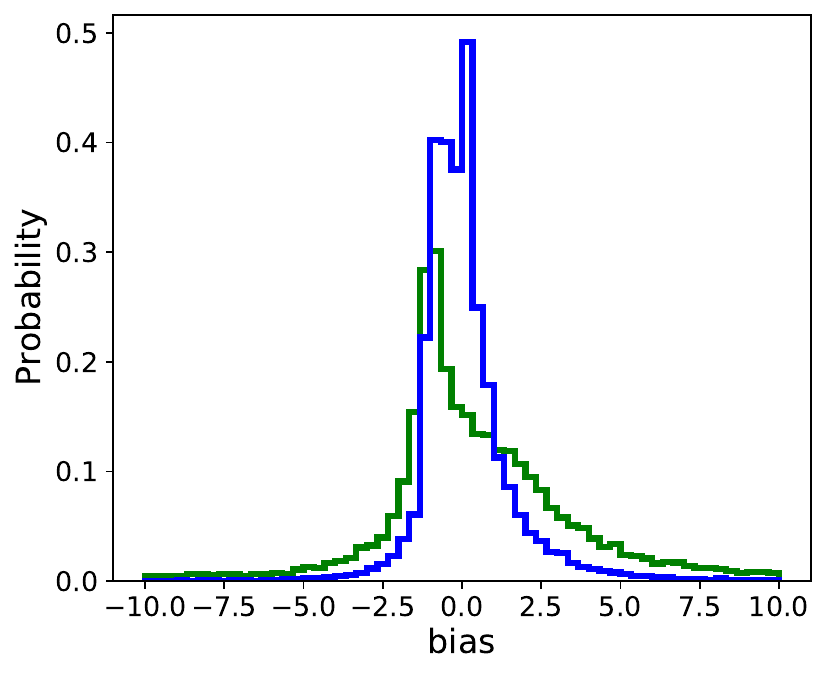}
\caption{Test with external simulations: The deep learning models precisely recover cluster masses for the sky maps with more realistic tSZ signal from the independent Magneticum hydrodynamical simulation.  
Shown is the histogram of bias for 30,000 CMB realizations using 140 tSZ profiles in the simulation (green contour).
The median bias of the distribution is -0.04 and standard error for 30,000 clusters is 0.04. 
The blue contour shows the distribution when Arnaud-based tSZ cluster profiles are used.
This shows that trained neural networks recover unbiased mass estimates at 4\% level for more realistic tSZ signal, even though the training is done with simplistic Arnaud profiles.
} 
\label{FIG:magneticum}
\end{figure}

\subsection{Testing model with external hydrodynamical simulations}
\label{sec:magneticum}

While we have shown machine learning works well on the symmetric Arnaud profile used for training, a reasonable question is how it will perform with more realistic (i.e.~complex) cluster profiles. 
We explore this question by running the trained network on  images drawn from clusters in the {\it Magneticum Pathfinder Simulation}\footnote{http://www.magneticum.org/} \citep[MPS][]{dolag16b, gupta17b, soergel18}. 

 The MPS is a large hydrodynamical simulation carried out as a counterpart to ongoing, multiwavelength surveys, and includes both the kSZ and tSZ effect. 
 We take cutouts of the kSZ and tSZ maps provided with the MPS simulations at the locations of $140$ galaxy clusters at $z=0.67$ and $z=0.73$ with 2$\times 10^{14}~\rm M_{\odot}$ $<M_{200\rm c}<$ 7$\times 10^{14}~\rm M_{\odot}$. 
The median mass of this sample is $M_{200\rm c} = 2.5 \times 10^{14}~\rm M_{\odot}$. 
The MPS cluster catalog lists masses in terms of the overdensity of 500 times the critical density of universe;  we convert this to $M_{200 \rm c}$ using a model of the concentration-mass relation \citep{diemer15}.
 As the MPS maps have a pixel size of $0.19^{\prime}$, we interpolate these maps to the match the $0.25^{\prime}$ pixelation used in this work. 
 We also convolve the SZ maps by the assumed  $1^{\prime}$ beam. 
The other signals in the mm-wave sky  (the lensed CMB, instrumental noise, radio and dusty galaxies) are generated as in Section~\ref{sec:SZE} and added to the SZ maps. 
To increase the statistical power of the sample, we create 30,000 realisations of lensed CMB, white noise and foregrounds for this sample of 140 SZ cluster profiles.

We pass the resulting maps to the two trained neural networks. 
Note that these networks have been trained with the symmetric Arnaud-based tSZ cluster profiles, not the MPS images. 
We plot the histogram of bias between the estimated and true masses in Figure~\ref{FIG:magneticum}. 
The median bias and standard error for 30,000 clusters is -0.04 and 0.04, respectively (green contour). 
For comparison, we also plot the distribution for 30,000 sky maps that have Arnaud-based tSZ cluster profiles with $M_{200\rm c} = (2, 2.5, 3) \times 10^{14}~\rm M_{\odot}$, finding median bias and standard error of -0.01 and 0.025 (blue contour). 
This shows that despite the networks being trained on symmetric Arnaud SZ profiles, the neural networks still produce unbiased mass estimates on realistic SZ profiles with 1 $\sigma$ limit for 30,000 clusters at 4\% level.

\begin{figure}
\centering
\includegraphics[width=9.4cm, scale=0.5]{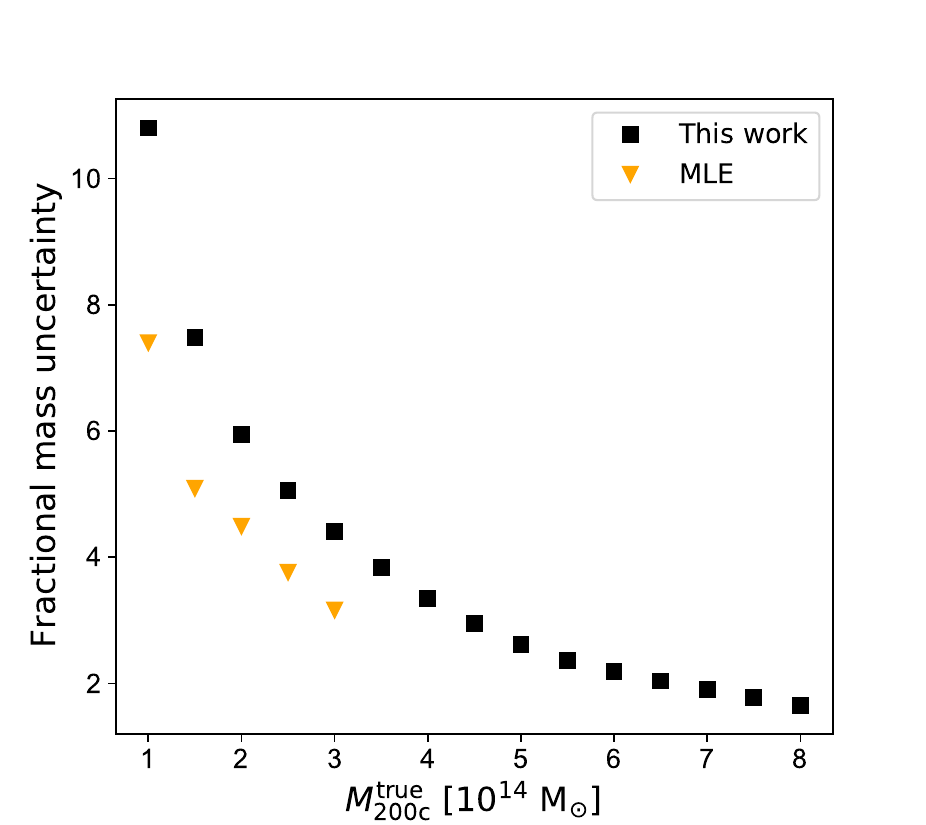}
\caption{Comparison with MLE: The fractional mass uncertainty on a single cluster as a function of input mass. 
The black squares show the $\Delta M_{\rm 200c}^{\rm est}/M_{\rm 200c}^{\rm est}$ from the current work.
The orange triangles  show the $\Delta M_{\rm 200c}^{\rm MLE}/M_{\rm 200c}^{\rm MLE}$ from the MLE using the noisy CMB maps with astrophysical backgrounds.
This shows that the fractional uncertainties from our deep learning method are factors of 1.3 to 1.45 times higher than those from the MLE. 
For instance, the fractional mass uncertainty for $2 \times 10^{14}~\rm M_{\odot}$ cluster is a factor 1.3 higher than the MLE limit.
} 
\label{FIG:deltaM_M}
\end{figure}
\subsection{Comparison with Mass Estimations from MLE}
\label{sec:MLE}
The MLE uses pixel to pixel correlations to fit lensed CMB templates to the observed microwave sky CMB maps.
The pixel to pixel covariance matrix is estimated using a set of simulated skies and is a function of cluster mass and redshift \citep[e.g.][]{baxter15}.
Figure~\ref{FIG:deltaM_M} shows the uncertainties in our mass estimations as a function of input mass (black squares).
The MLE uncertainties are estimated from the Gaussian realizations of CMB maps with 5 $\mu K$-arcmin white noise and astrophysical foregrounds. 
The tSZ signal is assumed to have been removed perfectly from these maps.
For uncertainties estimated with deep learning, we use test maps with Gaussian realizations of CMB, 5 $\mu K$-arcmin white noise, astrophysical foregrounds and tSZ signal. These maps are passed through already trained mResUNet-I model to reconstruct CMB-only maps with suppressed foregrounds and noise (see Section~\ref{sec:predictions1}). The reconstructed CMB only maps passed through the trained mResUNet-II model are used to estimate the uncertainties (see Section~\ref{sec:predictions2}).

With deep learning, we find per cluster $\Delta M_{\rm 200c}^{\rm est}/M_{\rm 200c}^{\rm est} =$ 10.8 and 1.6 for input cluster mass $M_{\rm 200c}^{\rm true}=10^{14}~\rm M_{\odot}$ and $8\times 10^{14}~\rm M_{\odot}$, respectively. 
We compare the deep learning numbers to the MLE uncertainties for $10^{14}~\rm M_{\odot} - 3\times 10^{14}~\rm M_{\odot}$ galaxy clusters with 5 $\rm \mu K$-arcmin noise with foregrounds (orange triangles).
Note that, the MLE uncertainty is estimated for $10^5$ clusters and the deep learning errors are for $10^4$ clusters (except for cluster mass $2\times 10^{14}~\rm M_{\odot}$  for which $10^5$ clusters are used to get deep learning errors), we scale them for one cluster for this comparison in Figure~\ref{FIG:deltaM_M}.
We find that the fractional mass uncertainties from the deep learning method are factors of 1.3 to 1.45 higher than MLE estimates in the compared mass range. For instance, the $\Delta M_{\rm 200c}^{\rm est}/M_{\rm 200c}^{\rm est} =$ 5.9 for $2\times 10^{14}~\rm M_{\odot}$ which is a factor 1.3 higher than the MLE limit.

The above comparisons show that the trained deep learning models put competitive constraints on the cluster masses.
Note that we do not use the reconstructed CMB-only maps with suppressed foregrounds and noise to derive the MLE mass uncertainties.
This is mainly due to the non-Gaussian nature of the residual noise in these maps (see Fig.~\ref{FIG:T_residual}).
The application of MLE to the reconstructed CMB-only maps will require new developments and we plan to modify the MLE in our future work.

\section{Conclusions}
\label{sec:conclusions}
We demonstrate for the first time a two-stage deep learning algorithm that first extracts CMB maps with foreground and noise suppressed, and second estimates the mass of galaxy clusters from their gravitational lensing imprint upon the CMB map. 

The CMB sky maps include Gaussian realizations of CMB and astrophysical foregrounds, cluster's own tSZ signal with 20\% intrinsic scatter (a foreground in this case), 5 $\rm \mu K$-arcmin instrumental white noise and 1$^{\prime}$ beam smoothing.
We train and validate the mResUNet-I network with 400 and 200 sky maps, respectively, for each of the eight cluster masses with  $M_{200\rm c} =$ (1, 2, 3, 4, 5, 6, 7, 8)$\times 10^{14}~\rm M_{\odot}$.
The mResUNet-II network is trained (400) and validated (200) using CMB maps with $M_{200\rm c} =~\pm$ (1, 2, 3, 4, 5, 6, 7, 8, 9, 10, 20...90, 100, 200...500)$\times 10^{14}~\rm M_{\odot}$, having white noise randomly drawn from a uniform distribution between 0 to $5~\rm \mu K$-arcmin level.
A test set of $10^4$ sky maps is used to first reconstruct CMB-only maps with suppressed foregrounds and noise using the trained mResUNet-I model and then to estimate the underlying cluster mass with the trained mResUNet-II model.
We find that the trained models recover the unbiased estimates of input mass.
For instance, the bias for cluster mass $4 \times 10^{14}~\rm M_{\odot}$ is $< 1.5\%$ with a 1 $\sigma$  limit for $10^4$ clusters at $3.5\%$ level.
As the uncertainty is larger for lower mass clusters, we also confirm the non-detection for a lower mass cluster with $M_{200 \rm c}^{\rm true}= 2 \times 10^{14}~\rm M_{\odot}$ using $10^5$ realizations, finding $1\%$ bias with a 2 $\sigma$ upper limit at 3.5\% level. 

As the cluster's own SZ signal acts as the brightest foreground for the CMB lensing signal, we test our trained models on more realistic SZ signal from the external hydrodynamical simulations.
While the mResUNet-I model is trained on simplified tSZ profiles (spherically symmetric Arnaud profiles), the trained model performs well when provided CMB images with more realistic SZ profiles. We demonstrate this by taking 140 galaxy cluster SZ cutouts from the light cones of the Magneticum hydrodynamical simulation at $z=0.67$ and $z=0.73$ with 2$\times 10^{14}~\rm M_{\odot}$ $<M_{200\rm c}<$ 7$\times 10^{14}~\rm M_{\odot}$. 
These cutouts include more complex tSZ structure from the cluster itself, kSZ signal from bulk motion of cluster, as well as the added tSZ contributions from other objects along nearby lines of sight.
We randomly add 30,000 realizations of CMB, noise and foregrounds to these SZ profiles and pass them through the trained deep learning networks.
Our trained mResUNet-I and mResUNet-II models recover the unbiased estimates of the true masses of the clusters with 1 $\sigma$ upper limit at 4\% level.

We compare the mass uncertainties from the deep learning models with those estimated from the MLE.
We find that the fractional mass uncertainties from the deep learning method are factors of 1.3 to 1.45 higher than MLE estimates in the compared mass range.
For instance, the per cluster $\Delta M_{\rm 200c}^{\rm est}/M_{\rm 200c}^{\rm est} =$ 5.9 for $2\times 10^{14}~\rm M_{\odot}$ which is a factor 1.3 higher than the MLE limit.

In our future work, we plan to apply this two-stage deep learning approach to estimate galaxy cluster masses in real CMB data. 
Presuming no insurmountable challenges appear, deep learning would be a valuable tool for determining the masses of high-redshift galaxy clusters in ongoing and upcoming CMB surveys \citep[e.g. SPT-3G, AdvancedACT, Simons Observatory, CMB-S4][]{benson14, henderson16, simons19, cmbs4-19}. 
Accurate mass estimates across the full range of redshifts will be essential to fully utilizing the large, $>10^5$ galaxy cluster samples expected from CMB, X-ray and optical surveys  \citep[e.g. {\it eROSITA},  LSST, Euclid][]{ predehl10, lsst09, laurejis11}.

\begin{acknowledgements}

We acknowledge support from the Australian Research Council's Discovery Projects scheme (DP150103208).
NG acknowledges support from CSIRO’s Machine Learning and Artificial Intelligence Future Science Platform.
This research uses resources of the National Energy Research Scientific Computing Center (NERSC).
We thank Srinivasan Raghunathan, Sanjay Patil, Brian Nord, Jo\~{a}o Caldeira and Federico Bianchini for their helpful feedback.

\end{acknowledgements}

\bibliographystyle{aasjournal}
\bibliography{CNN_clusters}

\end{document}